# Single-periodic-film optical bandpass filter

MANOJ NIRAULA, JAE WOONG YOON, AND ROBERT MAGNUSSON

*Department of Electrical Engineering, University of Texas at Arlington, Box 19016, Arlington, TX 76109, USA*

**Resonant periodic surfaces and films enable new functionalities with wide applicability in practical optical systems. Their material sparsity, ease of fabrication, and minimal interface count provide environmental and thermal stability and robustness in applications. Here we report an experimental bandpass filter fashioned in a single patterned layer on a substrate. Its performance corresponds to bandpass filters requiring ~30 traditional thin-film layers as shown by an example. We demonstrate an ultra-narrow, high-efficiency bandpass filter with extremely wide, flat, and low sidebands. This class of devices is designed with rigorous solutions of Maxwell's equations while engaging the physical principles of resonant waveguide gratings. The proposed technology is integration-friendly and opens doors for further development in various disciplines and spectral regions where thin-film solutions are traditionally applied.**

Thin-film optics is a mature technological area. Multilayer dielectric films are widely applied to implement metal-free and thus low-loss filters, polarizers, and reflectors for incorporation in common optical systems [1]. These devices typically consist of stacks of homogeneous layers deposited with precise thicknesses and tight control of index of refraction and absorption. A large number of layers, perhaps ~10–100, may be needed to create the spectral, polarization, and angular attributes required for a particular application. These optical devices operate on the basis of multiple reflections between the interfaces incorporated in a layered stack. In particular, periodic quarter-wave layer systems provide classical high reflectors for bulk laser cavities as well as integrated distributed Bragg reflectors for vertical cavity lasers [2]. Embedding a half-wave spacer layer between two reflector film stacks enables formation of a passband centered at the design wavelength [1]. This passband is the functional basis for commercial bandpass filters operating in the visible and near-infrared spectral domains with applications in telecommunications, laser-line filtering, astronomical observations, and spectroscopic and analytical instrumentation.

Resonant thin-film subwavelength gratings enable versatile optical properties for reflectors [3-5], spectral filters [6,7], and beam-transforming elements [8]. On account of their structural simplicity and on-chip integration compatibility, it is of interest to develop high-performance bandpass filters with flat, low sidebands and narrow, high-efficiency passbands. Previous related work has mostly focused on theoretical design and explanation of the underlying physics [9-11]. In an early proposal, quarter-wave Bragg stacks are used to generate broad stopbands accommodating narrow transmission peaks induced by resonant subwavelength gratings [9]. Practical designs with a single, or a few, thin-film layers have been found using inverse numerical optimization methods [10,11]. In these advanced designs, multiple leaky resonant modes with low-$Q$ and high-$Q$ resonance properties are forced to cooperate in a single nanopatterned layer to generate a desired spectral response [12]. A recent theoretical study in this context shows that near-unity peak efficiency and arbitrarily-narrow linewidth at a desired spectral location are attainable with partially-etched single-layer grating architectures even at normal incidence [13].

Experimental demonstration of resonant bandpass, or transmission, filters has been attempted by only few groups [14-16]. Kanamori *et al.* fabricated color filters with efficiencies in the range of ~55-70% but with very wide bandwidths ~100 nm and without suppression of the sidebands [14]. Foley *et al.* reported transmission resonance with passband efficiency ~35% and a broad peak bandwidth exceeding 500 nm in the mid-infrared domain; these devices operated only under off-normal incidence [15]. In another experimental demonstration, Amin *et al.* provided a Rayleigh-anomaly-assisted transmission filter with ~50% efficiency in the near-infrared domain; however, this design lacks the flat, low stopbands and well-defined narrow passband that are critical in practice [16].

Here, we report successful design, fabrication, and measurement of high-performance bandpass filters in the telecommunications band using single-layer resonant Si gratings. We obtain an ultra-narrow passband with peak efficiency ~72% and linewidth ~0.48 nm with attendant 100-nm-wide flat stopbands that lie below 1% in transmittance. These performance parameters satisfy requirements pertinent to wavelength-division multiplexing under the International Telecommunication Union standards [17]. The obtained passband is supported by an ultra-high-quality guided-mode resonance whose operating wavelength is adjustable by tuning the angle of incidence of input beam. Therefore, our results clearly demonstrate the feasibility of the proposed device class for bandpass filtering applications.

We apply a single-layer resonant thin-film grating structure on a silicon-on-quartz (SOQ) platform as shown in Fig. 1(a) in comparison with a classical multilayer structure in Fig. 1(b). The proposed single-layer structure has crystalline silicon (c-

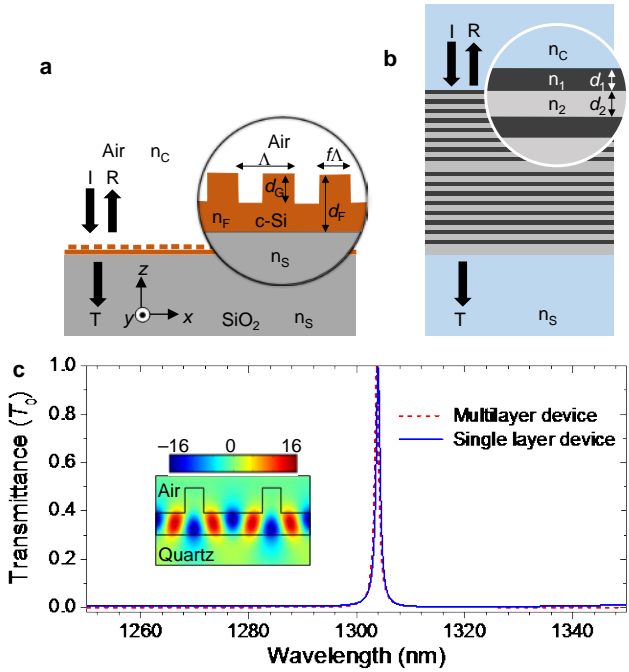

**Fig. 1.** Performance of a single-layer resonant bandpass filter in comparison with a traditional multilayer device. (a) Structure of a nanopatterned single-layer device. The device parameters are: refractive indices $n_C$ = 1.0 (Air), $n_F$ = 3.48 (Si), $n_S$ = 1.51 (SiO$_2$); film thicknesses $d_F$ = 520 nm and $d_G$ = 273 nm; period $\Lambda$ = 860 nm; and fill factor $f$ = 0.24. (b) Structure of a 32 layer bandpass filter with indices $n_1$ = 2.1, $n_2$ = 1.4 and $n_C$ = $n_S$ = 1.51. The corresponding quarter-wave thicknesses are $d_1$ = 155 nm and $d_2$ = 233 nm. Layers 16 and 32 from the top have a half-wave thickness of 466 nm. (c) Calculated zero-order transmittance ($T_0$) of the nanopatterned single-layer device in comparison with the multilayer device. We assume TE-polarized light at normal incidence. The inset in (c) shows the electric field distribution at the peak wavelength $\lambda$ = 1304 nm. The color scale is normalized by the incident electric-field amplitude.

Si) film of thickness $d_F$ on a quartz substrate with refractive index $n_S$ = 1.51. The top c-Si film is in the *x-y* plane and contains a one-dimensional grating with period $\Lambda$. The grating grooves lie along the *y*-axis. The width and height of a grating ridge is $f\Lambda$ and $d_G$, respectively. We optimize these parameters using the particle swarm optimization (PSO) method [11] to generate the desired bandpass filtering functionality under normally-incident transverse-electric (TE) polarized light in the telecommunications band around vacuum wavelength of 1300 nm. TE-polarized input light has its electric-field vector along the grating-ridge axis (*y*-axis). The rigorous coupled-wave analysis (RCWA) [18] is used as a forward kernel in the parametric optimization code. We model the c-Si film with its refractive index 3.48 which is experimentally obtained by ellipsometry. We note that the extinction coefficient of the c-Si is ~$10^{-6}$ and is therefore negligible in the numerical calculations.

Parametric optimization using the PSO method yields the zero-order transmittance ($T_0$) spectrum shown in Fig. 1(c). The optimized parameters are given in the figure caption. The $T_0$ spectrum has a passband with peak efficiency 99.8% and full-width at half-maximum (FWHM) linewidth $\Delta\lambda$ = 0.9 nm centered at $\lambda$ = 1304 nm. The normalized electric field ($E_y$) distribution at $\lambda$ = 1304 nm shown in the inset of Fig. 1(c) confirms that resonant coupling to the fundamental mode (TE$_0$) is responsible for the formation of the sharp transmission peak. The high-intensity fields associated with the standing TE$_0$ mode excitation are strongly confined in the homogeneous portion of the c-Si layer. Thus, the device performance is robust against minor parametric variations in the top patterned region [13]. The flat stopbands have $T_0 \leq 1\%$ over a 100-nm-wide wavelength range spanning from 1250 nm to 1350 nm. This remarkable stopband formation via the single nanopatterned c-Si layer is supported by a *resonant* broadband reflection effect and not a *thin-film interference* effect as explained in a recent study [13]. There, Niraula et al. show that the partially-etched grating device operates under a combination of two resonant modes: a low-*Q* TE$_1$ mode that forms the wide, flat stopband and a high-*Q* TE$_0$ mode that forms the narrow passband within the wide stopband.

Comparatively, a classical multilayer architecture with nearly identical performance requires 32 layers with typical refractive indices. It consists of a 31-layer Bragg stack and one embedded half-wave defect layer as illustrated in Fig. 1(b) with its length scale identical to that in Fig. 1(a); see the figure caption for the parameters of this multilayer device. Importantly, the single-layer grating architecture is advantageous over the classical multilayer structure in many practical aspects. For example, an array of different bandpass filters can easily be integrated on a single substrate. Such a filter array has potential applications in wavelength division multiplexing systems and compact spectrum analyzers without a wavelength-dispersive element. We also note that a variety of dielectric materials and structures is available for the proposed architecture in the visible, near-infrared, and longer wavelengths [10] where the deposition of quarter or half-wave

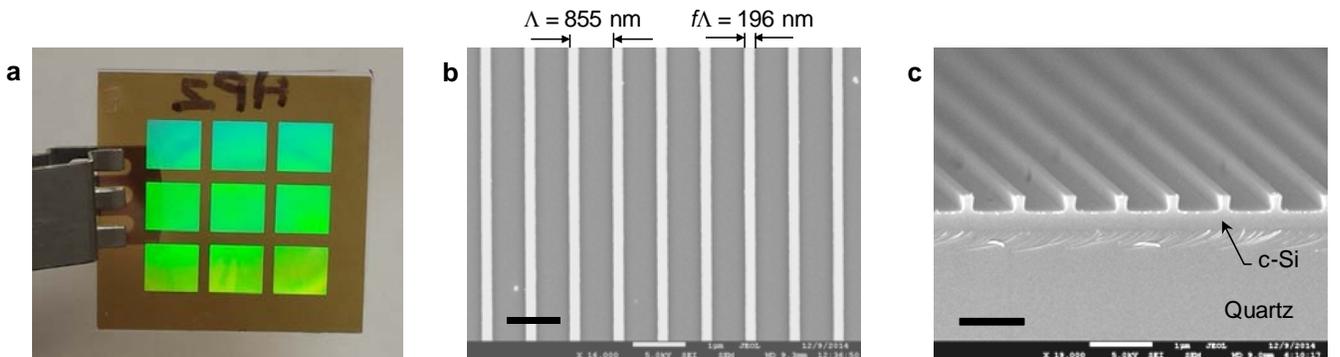

**Fig. 2.** Fabricated bandpass filter. (a) Photograph of nine fabricated devices on a 1×1 inch$^2$ silicon-on-quartz wafer. Each filter measures 5×5 mm$^2$. (b) Top-view and (c) cross-sectional scanning-electron micrographs of a representative device. Bar scales are 1 μm in both (b) and (c).

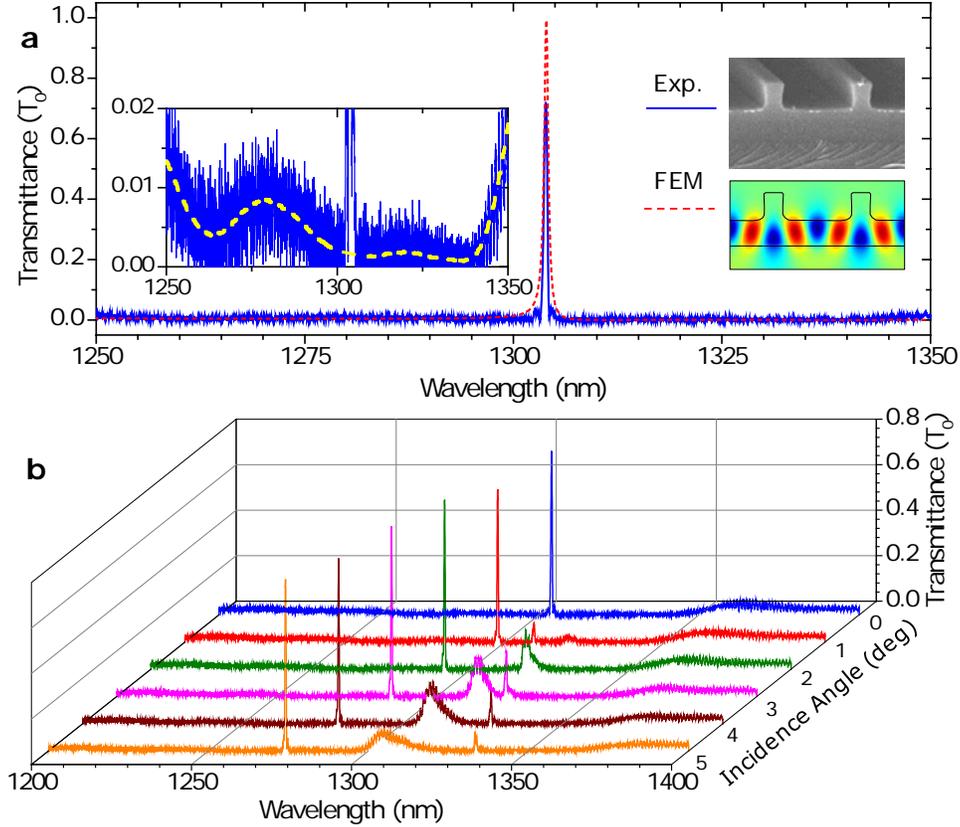

**Fig. 3.** Experimental performance of a fabricated device. (a) Measured transmittance ($T_0$) spectrum in comparison with the FEM simulation. In the FEM simulation, we take the exact experimental structure as indicated with the plot legend on the right. The inset graph shows a magnified view of the measured filter spectrum before (blue solid) and after (yellow dashed) numerical noise rejection. Note that the source noise level is ~1% in transmittance. (b) Measured $T_0$ spectra (raw data) for several small angles of incidence as indicated in the plot.

multilayers is particularly tedious or impractical.

We experimentally demonstrate the proposed bandpass filter and its performance. In the fabrication, we use an SOQ wafer with 520-nm-thick c-Si on a quartz substrate. The fabrication processes include ultraviolet holographic lithography to create a periodic photoresist mask, reactive-ion etching through the c-Si layer with a $CHF_3+SF_6$ gas mixture, and $O_2$ ashing for residual photoresist mask removal. Figure 2(a) shows nine fabricated devices on a 1×1 inch$^2$ SOQ wafer. Using scanning electron microscopy (SEM), we characterize the geometrical parameters of the fabricated device. Top-view and cross-sectional SEM images of a device are shown in Figs. 2(b) and 2(c). The estimated parameters for the device with the best performance are period $\Lambda$ = 855 nm, grating-ridge width $f\Lambda$ = 196 nm, and grating depth $d_G$ = 272 nm.

For spectral measurements, we use a Koheras SuperK Compact super-continuum light source and a Yokogawa AQ6375 near-infrared optical spectrum analyzer (OSA). The input beam is highly collimated and TE-polarized with ~1 mm spot size. A single-mode optical fiber is used to transport the transmitted signal. $T_0$ is calculated as transmitted signal normalized by the input light intensity. An adequate sampling resolution of 0.02 nm is maintained in the OSA. The measured $T_0$ spectrum at normal incidence $\theta$ = 0 is shown in Fig. 3(a). The narrow passband peak at $\lambda$ = 1304 nm has an efficiency of 72.1% and a FWHM linewidth of $\Delta\lambda$ = 0.48 nm. The resonance quality factor ($Q$-factor) is $\lambda/\Delta\lambda \approx 2.7\times10^3$. We note that higher $Q$-factors are theoretically attainable with this bandpass filtering principle [13]. The wide stopbands constituting the filter sidebands covering a 100-nm-wide wavelength region from 1250 nm to 1350 nm have $T_0$ < 1% as shown in the inset of Fig. 3(a). The theoretical spectrum in Fig. 3(a) is calculated using the Finite Element Method (FEM) and quantitatively agrees with the experimental spectrum. In the FEM calculation, we apply the exact device cross-section as indicated by the plot legend on the right in Fig. 3(a). We believe the degradation in the experimental peak efficiency is mainly due to scattering and absorption losses due to particle contamination of the SOQ wafer as visible in Fig. 2(b). This contamination does not significantly impact the stopbands as they are a result of a low-$Q$ resonance. These losses can be overcome with device fabrication in a cleanroom facility with a low particle count. Note that the reflection from the quartz substrate results in a ~4% drop in peak efficiency. An angle-dependent $T_0$ spectrum is shown in Fig. 3(b). Here, for every 1° increment in $\theta$, the location of the transmission peak shifts by ~6 nm to a shorter wavelength while maintaining the high peak efficiency, narrow linewidth, and low stopbands. Therefore, this device permits continuous spectral tuning of the passband under angular adjustment.

In conclusion, we provide the first experimental demonstration of an ultra-narrow bandpass filter in the near-infrared spectral domain using all-dielectric resonant gratings. We design, fabricate, and test nanostructured single-layer bandpass filters performing with a high-efficiency, sub-nanometer-wide passband and 100-nm-wide stopbands. The proposed device class is compatible with standard nanolithography processes and applicable to the visible/infrared/THz and longer wavelength domains. Recalling extraordinary optical transmission (EOT) through plasmonic nanoaperture arrays, nanostructured metallic bandpass filters have been fabricated [19]. However, the unavoidable inherent loss of metals results in low passband efficiencies and broad linewidths of EOT filters [19-21]. Therefore, further developing the device class proposed here is important for a host of applications. For example, this technology can be basis for wavelength-division demultiplexers, compact arrayed high-resolution spectrometers, hyper-spectral imaging, and Raman-scattering-based molecular-composition analyzers.

In future work, the results presented here can be improved by implementing crossed-grating designs for unpolarized light filtering. Recently, Shokooh-Saremi *et al.* proposed polarization-independent two-dimensional (2D) grating broadband mirrors designed using the PSO algorithm [22]. Hence, PSO is an important tool also for designing polarization-independent bandpass filters. Experimental issues, including parametric fabrication tolerances, associated with bandpass filters enabled by 2D grating profiles must also be considered.

**Funding.** The research leading to these results was supported in part by the Texas Instruments Distinguished University Chair in Nanoelectronics endowment.

**Acknowledgement.** The authors thank Shin-Etsu Chemical, Co., Ltd., Japan, for providing the SOQ wafers used in this research.

**References**
1. H. A. Macleod, Thin-Film Optical Filters (McGraw-Hill, 1989).
2. B. E. A. Saleh, M. C. Teich, Fundamentals of Photonics (Wiley, 2007).
3. Y. Fink, *et al.*, Science **282,** 1679 (1998).
4. C. F. R. Mateus, M. C. Y. Huang, Y. Deng, A. R. Neureuther, C. J. Chang-Hasnain, IEEE Photon. Technol. Lett. **16,** 518 (2004).
5. R. Magnusson, Opt. Lett. **39,** 4337 (2014).
6. S. Peng, G. M. Morris, J. Opt. Soc. Am. A **13,** 993 (1996).
7. D. W. Peters, *et al*., Opt. Lett. **35,** 3201 (2010).
8. D. Fattal, J. Li, Z. Peng, M. Fiorentino, R. G. Beausoleil, Nat. Photon. **4,** 466 (2010).
9. S. Tibuleac, R. Magnusson, IEEE Phot. Tech. Lett. **9,** 464 (1997).
10. S. Tibuleac, R. Magnusson, Opt. Lett. **26,** 584 (2001).
11. M. Shokooh-Saremi, R. Magnusson, Opt. Lett. **32,** 894 (2007).
12. Y. Ding, R. Magnusson, Opt. Lett. **29,** 1135 (2004).
13. M. Niraula, J. W. Yoon, R. Magnusson, Opt. Exp. **22,** 25817 (2014).
14. Y. Kanamori, M. Shimono, K. Hane, IEEE Phot. Tech. Lett. **18,** 2126 (2006).
15. J. M. Foley, S. M. Young, J. D. Phillips, Appl. Phys. Lett. **103,** 071107 (2013).
16. M. S. Amin, J. W. Yoon, R. Magnusson, Appl. Phys. Lett. **103,** 131106 (2013).
17. International Telecommunication Union, Spectral grids for WDM applications: DWDM frequency grid. (ITU-T G.694.1, 2012; http://www.itu.int/rec/T-REC-G.694.1-201202-I/en).
18. M. G. Moharam, D. A. Pommet, E. B. Grann, T. K. Gaylord, J. Opt. Soc. Am. A **12,** 1077 (1995).
19. T. W. Ebbesen, H. J. Lezec, T. T. Ghaemi, P. A. Wolff, Nature **391,** 667 (1998).
20. U. Schröter, D. Heitmann, Phys. Rev. B **58,** 15419 (1998).
21. E. Sakat, *et al.*, Opt. Lett. **36,** 3054 (2011).
22. M. Shokooh-Saremi, R. Magnusson, Opt. Lett. **39**, 6958 (2014).

**DETAILED REFERENCES**

1. H. A. Macleod, *Thin-Film Optical Filters* (McGraw-Hill, New York, ed. 3, 1989).
2. B. E. A. Saleh, M. C. Teich, *Fundamentals of Photonics* (Wiley, New York, ed. 2, 2007).
3. Y. Fink, *et al.*, A dielectric omnidirectional reflector. *Science* **282,** 1679-1682 (1998).
4. C. F. R. Mateus, M. C. Y. Huang, Y. Deng, A. R. Neureuther, C. J. Chang-Hasnain, Ultrabroadband mirror using low-index cladded subwavelength grating. *IEEE Photon. Technol. Lett.* **16,** 518–520 (2004).
5. R. Magnusson, Wideband reflectors with zero-contrast gratings. *Opt. Lett.* **39,** 4337–4340 (2014).
6. S. Peng, G. M. Morris, Resonant scattering from two-dimensional gratings. *J. Opt. Soc. Am. A* **13,** 993–1005 (1996).
7. D. W. Peters, *et al.*, Demonstration of polarization-independent resonant subwavelength grating filter arrays. *Opt. Lett.* **35,** 3201-3203 (2010).
8. D. Fattal, J. Li, Z. Peng, M. Fiorentino, R. G. Beausoleil, Flat dielectric grating reflectors with focusing abilities. *Nat. Photon.* **4,** 466–470 (2010).
9. S. Tibuleac, R. Magnusson, Diffractive narrow-band transmission filters based on guided-mode resonance effects in thin-film multilayers. *IEEE Phot. Tech. Lett.* **9,** 464-466 (1997).
10. S. Tibuleac, R. Magnusson, Narrow-linewidth bandpass filters with diffractive thin-film layers. *Opt. Lett.* **26,** 584–586 (2001).
11. M. Shokooh-Saremi, R. Magnusson, Particle swarm optimization and its application to the design of diffraction grating filters. *Opt. Lett.* **32,** 894-896 (2007).
12. Y. Ding, R. Magnusson, Doubly resonant single-layer bandpass optical filters. *Opt. Lett.* **29,** 1135-1137 (2004).
13. M. Niraula, J. W. Yoon, R. Magnusson, Mode-coupling mechanisms of resonant transmission filters. *Opt. Exp.* **22,** 25817-25829 (2014).
14. Y. Kanamori, M. Shimono, K. Hane, Fabrication of transmission color filters using silicon subwavelength gratings on quartz substrates. *IEEE Phot. Tech. Lett.* **18,** 2126-2128 (2006).
15. J. M. Foley, S. M. Young, J. D. Phillips, Narrowband mid-infrared transmission filtering of a single layer dielectric grating. *Appl. Phys. Lett.* **103,** 071107 (2013).
16. M. S. Amin, J. W. Yoon, R. Magnusson, Optical transmission filters with coexisting guided-mode resonance and Rayleigh anomaly. *Appl. Phys. Lett.* **103,** 131106 (2013).
17. International Telecommunication Union, *Spectral grids for WDM applications: DWDM frequency grid.* (ITU-T G.694.1, 2012; http://www.itu.int/rec/T-REC-G.694.1-201202-I/en).
18. M. G. Moharam, D. A. Pommet, E. B. Grann, T. K. Gaylord, Stable implementation of the rigorous coupled-wave analysis for surface-relief gratings: enhanced transmittance matrix approach. *J. Opt. Soc. Am. A* **12,** 1077-1086 (1995).
19. T. W. Ebbesen, H. J. Lezec, T. T. Ghaemi, P. A. Wolff, Extraordinary optical transmission through sub-wavelength hole arrays. *Nature* **391,** 667-669 (1998).
20. U. Schröter, D. Heitmann, Surface-plasmon-enhanced transmission through metallic gratings. *Phys. Rev. B* **58,** 15419-15421 (1998).
21. E. Sakat, *et al.*, Guided mode resonance in subwavelength metallodielectric free-standing grating for bandpass filtering. *Opt. Lett.* **36**, 3054–3056 (2011).
22. M. Shokooh-Saremi, R. Magnusson, Properties of two-dimensional resonant reflectors with zero-contrast gratings. *Opt. Lett*. **39,** 6958-6961 (2014).